# Strong interface-induced spin-orbit coupling in graphene on WS$_2$

Zhe Wang[1], Dong-Keun Ki[1], Hua Chen[2], Helmuth Berger[3], Allan H. MacDonald[2] and Alberto F. Morpurgo[1*]

[1]*Department of Quantum Matter Physics (DQMP) and Group of Applied Physics (GAP), University of Geneva, 24 Quai Ernest-Ansermet, CH1211 Genève 4, Switzerland*

[2]*Department of Physics, The University of Texas at Austin, Austin, Texas 78712, USA*

[3]*Institut de Physique de la Matière Complexe, Ecole Polytechnique Federale de Lausanne, CH-1015 Lausanne, Switzerland*

[*]e-mail: Alberto.Morpurgo@unige.ch

**Interfacial interactions allow the electronic properties of graphene to be modified, as recently demonstrated by the appearance of satellite Dirac cones in the band structure of graphene on hexagonal boron nitride (hBN) substrates. Ongoing research strives to explore interfacial interactions in a broader class of materials in order to engineer targeted electronic properties. Here we show that at an interface with a tungsten disulfide (WS$_2$) substrate, the strength of the spin-orbit interaction (SOI) in graphene is very strongly enhanced. The induced SOI leads to a pronounced low-temperature weak anti-localization (WAL) effect, from which we determine the spin-relaxation time ($\tau_{so}$). We find that $\tau_{so}$ in graphene is two-to-three orders of magnitude smaller on WS$_2$ than on SiO$_2$ or hBN, and that it is comparable to the intervalley scattering time ($\tau_{iv}$). To interpret our findings we have performed first-principle electronic structure calculations, which both confirm that carriers in graphene-on-WS$_2$ experience a strong SOI and allow us to extract a spin-dependent low-energy effective Hamiltonian. Our analysis further shows that the use of WS$_2$ substrates opens a possible new route to access topological states of matter in graphene-based systems.**

Because of the Dirac nature of its charge carriers and the presence of two valleys, graphene is a two-dimensional topological insulator[1,2]. Topological state characteristics have not been observed experimentally, because the strength of the SOI intrinsically present in graphene is too weak[3-5]. Various strategies to amplify the SOI strength have been proposed theoretically[6-8] or explored experimentally[8-11]. However, increasing the SOI strength in graphene without drastically affecting other basic aspects of its electronic structure, or the material quality, is proving extremely difficult[8-11]. Here, we explore whether it is possible to induce strong SOI while preserving the quality of graphene, by exploiting interfacial interactions[12-14] at an atomically sharp interface between graphene and a semiconducting WS$_2$ crystalline substrate.

Many semiconducting transition metal dichalcogenides (TMDs), such as WS$_2$, are ideal substrates for graphene. Like hexagonal boron nitride (hBN)[15], they are atomically flat and chemically inert, which is key to preserving high-quality transport properties (mobility values as high as $\mu \sim$ 50,000-60,000 cm$^2$/Vs have been recently reported for graphene-on-WS$_2$)[16]. TMD crystals consist of a stack of monolayers having a hexagonal lattice that –like graphene– leads to the presence of two valleys in their electronic structure at the K and K' point of the Brillouin zone[17]. The SOI in WS$_2$ is extremely strong –several hundreds of millivolts in the valence bands and several tens of millivolts in the conduction band– and in monolayers it pins spin to valley[18-20]. The spins of states near the band edges point in one direction in one of the valleys and in the opposite direction in the other, a behavior resembling the one expected theoretically in disorder free graphene[1,2]. The ability of this substrate material to induce a strong SOI in graphene –as well as the nature of the induced SOI– are, therefore, important topics that have attracted recent attention[10]. Here, we address these issues by combining a study of low-temperature quantum transport in graphene-on-WS$_2$ devices with *ab-initio* electronic structure calculations.

We start by characterizing the basic transport properties of the graphene-on-WS$_2$ devices used in our experiments. The devices are assembled in a multi-terminal Hall bar configuration, placed on a highly doped Si wafer which acts as a gate electrode, and coated with 280 nm SiO$_2$ (see the upper inset of Fig. 1a for an optical microscope image, Fig. 1b for a schematic of the device structure, and the Methods section for the details of the device fabrication). We have realized several such devices. Here we present representative data from one of them (similar data from another device can be found in the Supplementary Information). Fig. 1a shows that upon ramping up the gate voltage from $V_g$ = -40 V, the conductivity $\sigma$ of graphene-on-WS$_2$ decreases linearly until $V_g \sim$ 8 V, after which it saturates. Saturation occurs because, for $V_g >$ 8 V, electrons are accumulated at the interface between the SiO$_2$ and the WS$_2$ crystal, screening the effect of the gate on the graphene layer on top (Fig. 1b)[21]. (Because the mobility of charge carriers in WS$_2$ is much smaller than in graphene[22], carriers at the WS$_2$/SiO$_2$ interface give a negligible contribution to transport.) Sweeping the gate voltage down for $V_g <$ 8 V, on the contrary, results in an increase (Fig. 1b) of carrier (hole) density in graphene and the conductivity increases. In our devices, therefore, the position of the Fermi level can be gate shifted in the graphene valence band, but accumulation of electrons at the SiO$_2$/WS$_2$ interface frustrates access to the conduction band. In the following we therefore study only hole transport through graphene. To illustrate this conclusion –and to start assessing the device quality– Fig. 1c shows that the half-integer quantum Hall effect characteristic of monolayer graphene[23] is clearly observed for different values of $V_g$ between –40 and 8 V. The longitudinal resistance measured in this $V_g$ range, and plotted versus filling factor $\nu$ and magnetic field $B$ (bottom inset of Fig. 1a), confirms this result. We estimate the carrier mobility from $\sigma$ using the hole density $n$ extracted from the (classical) Hall effect and by looking at the slope $d\sigma/dV_g$, and obtain in both cases $\mu \sim 13\,000$ cm$^2$/Vs at $T = 4.2$ K. Since no effort has yet been put into optimizing the fabrication process, these values confirm the very good quality of graphene-on-WS$_2$ devices found in earlier work[16].

To demonstrate the presence of SOI in our devices, we probe weak antilocalization (WAL), which usually manifests itself as a characteristic sharp magnetoconductance (MC) peak at $B = 0$ T[24,25]. In small, fully phase coherent devices like ours, however, WAL is eclipsed by conductance fluctuations originating from the random interference of electron waves[25,26]. Indeed, in Fig. 2a, which shows the MC as a function of $V_g$ and $B$, an enhancement in conductance at $B = 0$ T is only faintly visible. No special feature at $B = 0$ T can be detected by looking at a single MC curve measured at a fixed value of $V_g$ (see, e.g., the top curve shown in Fig. 2b measured at $V_g = -25$ V). The random conductance fluctuations, whose reproducibility is shown in Fig. 2c, can be suppressed through an ensemble averaging procedure in which MC traces measured at different $V_g$ values are averaged[25]. The $V_g$ spacing should be chosen to shift the Fermi level by Thouless energy of the system. It is expected that the root-mean-square amplitude of the fluctuations decreases as $N^{1/2}$ ($N$ is number of uncorrelated MC traces used to calculate the average), eventually making the sharp conductance peak at $B = 0$ T due to WAL visible, if the strength of SOI is sufficient. This is indeed what the experiments show (see Figs. 2b and 2e).

We find that the WAL signal emerging from the ensemble average procedure is robust, and visible in the entire $V_g$ range investigated. Its amplitude grows upon lowering temperature $T$ (see Figs. 3a-3c), and reaches $\sim 0.5 \times e^2/h$ at the largest negative $V_g$ and $T = 250$ mK (Fig. 3c). The phenomenon is not observed in graphene on conventional substrates such as SiO$_2$[27-29] (or hBN[30] or GaAs[31]), where at sub-Kelvin temperatures only weak localization is measured. This remark is important because in graphene WAL can occur also in the absence of SOI, due only to the Dirac nature of its charge carriers[32-34]. The WAL originating from the Dirac nature of electrons, however, is unambiguously different from what we observe on WS$_2$ substrates: it is seen only for $T \sim 10$ K or higher, and has small amplitude, because its observation requires the phase coherence time $\tau_\phi$ to be shorter than the intervalley scattering time $\tau_{iv}$[28]. The observation of the low-temperature magnetoconductance shown in Fig. 3, therefore, represents a direct, unambiguous demonstration of the presence of SOI in graphene on a WS$_2$ substrate.

To analyze the MC data quantitatively, we use the theory of WAL in graphene that considers the effect of all possible symmetry-allowed SOI terms in graphene, and predicts the following dependence of the low-temperature MC[35]:

$$\Delta\sigma(B) = -\frac{e^2}{2\pi h}\left[F\left(\frac{\tau_B^{-1}}{\tau_\varphi^{-1}}\right) - F\left(\frac{\tau_B^{-1}}{\tau_\varphi^{-1}+2\tau_{asy}^{-1}}\right) - 2F\left(\frac{\tau_B^{-1}}{\tau_\varphi^{-1}+\tau_{so}^{-1}}\right)\right] \quad (1).$$

Here, $\tau_{asy}^{-1}$ is the rate of spin-relaxation uniquely due to the SOI terms that break the $z \to -z$ symmetry ($z$ being the direction normal to the graphene plane), $\tau_{so}^{-1}$ is the total spin-relaxation rate due to all SOI terms present, $\tau_B^{-1} = 4DeB/\hbar$ ($D$ is the carrier diffusion constant), and $F(x) = \ln(x) + \psi(1/2 + 1/x)$ with the digamma function $\psi(x)$. In fitting the data, we constrain all characteristic times to be independent of temperature, except for $\tau_\phi$, which increases upon lowering $T$, as physically expected in the $T$ range investigated. Eq. (1) holds in the limit $\tau_\phi \gg \tau_{iv}$, which is the one physically relevant at low $T$, and reproduces the experimental results quantitatively (solid lines in Figs. 3a-c). The analysis allows us to obtain the relevant characteristic times $\tau_{so}$, $\tau_{asy}$, and $\tau_\phi$, with a precision determined by the residual conductance fluctuations that are not perfectly removed by the ensemble averaging procedure. (These residual effects also cause $\sigma$ to be non-perfectly symmetric upon reversing $B$, because the conductivity is extracted from the conductance measured in a four-terminal configuration, which in fully phase coherent devices is in general not symmetric[25].) We conservatively estimate the error on the characteristic times to be approximately 50% in the worst case: although rather large as compared to what can be achieved in more established material systems, such an uncertainty is immaterial for all the considerations that will follow.

We find the spin-relaxation time to be $\tau_{so} \sim 2.5 \sim 5$ ps depending on the value of $V_g$ (see the black filled circles in Fig. 3d). Comparable values (within experimental uncertainties) have been obtained on the same devices from the analysis of measurements of non-local resistance generated by spin-Hall and inverse spin-Hall effect[36] (see red circles in the left panel of Fig. 3d and Supplementary Information for details). The latter technique was used recently in Refs. 9 and 10 to probe SOI in hydrogenated graphene and graphene on $WS_2$ in devices analogous to ours (see the Supplementary Information for a comparison). The value of $\tau_{asy}$ is approximately three times larger than $\tau_{so}$, consistent with its physical meaning; the phase coherence time $\tau_\phi \gg \tau_{so}$, as it must be since a large WAL signal is observed (see the right panel in Fig. 3d; $\tau_\phi$ decreases upon increasing $T$, as expected). This internal consistency of the hierarchy of characteristic times extracted from fitting the data with Eq. (1) supports the validity of our analysis. We conclude that in graphene at a $WS_2$ interface $\tau_{so}$ is 100-1000 times shorter than $\tau_{so}$ for pristine graphene on $SiO_2$[37,38] or hBN[39,40] (shown with open circles in Fig. 3d), and that such a large difference in strength must be due to substantially stronger SOI. In contrast to what has been reported in recent studies of graphene-on-$WS_2$ (see Ref. 10 and Supplementary Information), the larger strength persists throughout the entire gate voltage range investigated.

Determining the precise nature of the $WS_2$-induced SOI is not straightforward. A customary way to extract information is to identify the spin-relaxation mechanism by looking at how $\tau_{so}$ depends on $\tau$, the transport scattering time. Finding that $\tau_{so}$ increases with increasing $\tau$ points to the so-called Elliot-Yafet relaxation mechanism (spin-relaxation mediated by scattering at impurities)[41,42], whereas if $\tau_{so}$ decreases with increasing $\tau$, the Dyakonov-Perel mechanism (typical of systems with a strong band SOI) may be invoked[43]. In graphene on $SiO_2$ or hBN substrates, previous work has shown that neither scenario convincingly accounts for the observations[8], which has led to both phenomenological approaches to describe the experimental data[39], and to the theoretical proposal of new spin-relaxation mechanisms specific to graphene[44]. For graphene-on-$WS_2$, despite the much larger strength of SOI, the interpretation of spin-relaxation within the canonical schemes poses similar problems: $\tau_{so}$ decreases slightly upon increasing $\tau$ (see the left panel in Fig. 3d), ruling out Elliot-Yafet as a dominant relaxation mechanism, but the dependence is much weaker than the one predicted by Dyakonov and Perel, $\tau_{so} \propto 1/\tau$, so that the data are not satisfactorily described by this mechanism either. One interesting observation, however, can be made by comparing the spin-relaxation time $\tau_{so}$, with the intervalley scattering time $\tau_{iv}$ obtained from the analysis of weak localization in graphene on SiO$_2$[27,29], hBN[30], and GaAs[31]. Literature values of $\tau_{iv}$ are shown with empty triangles in Fig. 3d: they are surprisingly narrowly distributed –they all fall within a factor of 2– if we consider that experiments have been performed by different groups and that different substrate materials are used. The values of $\tau_{iv}$ match well (again, within a factor of 2 or better) $\tau_{so}$ obtained for graphene-on-WS$_2$. This close correspondence between two *a priori* unrelated quantities is remarkable: it strongly suggests that the microscopic processes responsible for scattering between the two valleys are the same processes that cause spin flip in graphene-on-WS$_2$.

To gain a better understanding of our experimental findings we have performed electronic structure calculations for a large number of crystal approximants to graphene-on-WS$_2$. Typical results are illustrated schematically in Fig. 4b, and the calculations are described in detail in the Supplementary Information. We find that states near the graphene Dirac point are always accurately described by effective Hamiltonians of the form

$$H = H_0 + \frac{\Delta}{2} \sigma_z + \frac{\lambda}{2} \tau_z s_z + \frac{\lambda_R}{2} \left( \tau_z \sigma_x s_y - \sigma_y s_x \right) \quad (2),$$

where **σ** is a Pauli matrix vector which acts on the sub lattice degree-of-freedom in graphene's Dirac continuum model Hamiltonian $H_0$, **s** is a Pauli matrix vector which acts on spin, and $\tau_z = \pm 1$ for K and K' valleys. All three substrate-induced interaction terms are time-reversal invariant and absent by inversion symmetry in isolated graphene sheets. Unlike the Hamiltonians that describe graphene on h-BN[45], $H$ in Eq. (2) is translationally invariant. The fact that such a model is able to describe graphene states is expected because the lattice constant difference between graphene and WS$_2$ is much larger than the lattice constant difference between graphene and h-BN. The moiré pattern period is correspondingly shorter and does not couple pristine graphene states within the WS$_2$ gap which are isolated near the graphene Brillouin-zone corners.

Numerical values for the substrate interaction parameters ($\Delta$, $\lambda$, and $\lambda_R$) depend on the supercell commensurability between graphene and WS$_2$ triangular lattices. This dependence demonstrates that even if each layer had a two-dimensional lattice that was free of defects, the continuum model of Eq. (2) should be supplemented by spin-dependent terms of a similar size which vary rapidly in space, can scatter graphene electrons between valleys, and will behave like random potentials in the realistic case of incommensurate structures. This random spin-dependent substrate interaction explains our finding that $\tau_{so}$ in graphene-on-WS$_2$ is comparable to $\tau_{iv}$. The largest supercells we considered had a 9:7 ratio between the WS$_2$ and graphene lattice constants, very nearly in perfect agreement with experiment. We find that the interaction parameters implied by this commensurability are independent of rigid relative translations between graphene and the substrate, providing further support for the translational invariance of Eq. (2). They are however sensitive to the separation between graphene and substrate layers, which has not yet been accurately determined experimentally, but should be near 3 Å like for graphene-on-hBN[46]. For such a separation our calculations give $\Delta \approx 0$ meV, $\lambda \approx 5$ meV, and $\lambda_R \approx 1$ meV. The two larger SOI parameters exceed the scale of the SOI in isolated graphene sheets by two to three orders of magnitude[3-5]. Consistently with the experimental results, therefore, the results of our calculations point to a strong enhancement of SOI.

The general form of the Hamiltonian in Eq. (2), which yields a gap at charge neutrality because the Rashba term couples conduction and valence band states (compare Figs. 4b and 4c), deserves comment. It can be shown (see Supplementary Information) that when the Fermi energy lies in this gap, the system is a topological insulator. Even though the energy scale of this topological insulating state is significantly larger (by one-to-two orders of magnitude) than that based on the intrinsic SOI in graphene originally proposed by Kane and Mele[2], its presence will be reflected in non-local transport experiments[47] only if charge inhomogeneity or other types of disorder are sufficiently weak. Finding a way to further increase the value of the parameters $\lambda$ and $\lambda_R$ experimentally is therefore particularly interesting. For example, experiments in which the Dirac point can be reached by gating and pressure is applied to decrease the separation between the graphene sheet and the underlying WS$_2$ crystal appear particularly promising.

The increase in spin-obit interaction strength induced in graphene by proximity with WS$_2$ that we observe for all values of gate voltage –and hence position of the Fermi energy– investigated is a large effect, close to two orders of magnitude. It directly shows the relevance of interfacial interactions. Demonstrating the ability to combine the control given by these interactions with the high electronic quality inferred from the experiments is a key result: although clearly more optimization is needed, finding that a drastic increase in SOI strength can be achieved without compromising the electronic quality of graphene offers the possibility to improve the system quality even further. To this end, strategies already exist that largely benefit from the expertise developed for graphene and hBN, such as encapsulating graphene between two WS$_2$ crystals[16], using different device assembly techniques to avoid contact between graphene and unwanted materials during fabrication[48], and optimizing the device cleaning procedures[49]. As indicated by both our experimental results and their theoretical modelling, these developments open a possible route to access experimentally topological states[50] in graphene, and will play an essential role in improving our understanding of the spin dynamics in graphene.

**Methods**

Thin WS$_2$ flakes were exfoliated onto a highly doped silicon substrate acting as a gate covered by a SiO$_2$, from high quality single crystals of WS$_2$ grown by chemical vapor transport method. These exfoliated WS$_2$ flakes were annealed at 200 degrees for 3 hour in an inert atmosphere, after which a specific WS$_2$ flake with atomically flat, and clean surface was identified through imaging with an atomic force microscope. Transfer of graphene onto the selected WS$_2$ flake was achieved by means of (by now) common techniques[15]. Similarly to the case of other artificial stacks of atomically thin crystals, bubbles were found to form after transfer of graphene (visible as black points in the upper inset of Fig. 1a)[30]. To minimize the effect of these bubbles on the transport experiments, graphene was etched into a Hall-bar geometry, after which electrodes consisting of a Ti/Au bilayer (10/70 nm) were defined by means of electron-beam lithography, electron-beam evaporation and lift-off. No post-annealing steps, or other cleaning processes, to further improve the device quality were performed on the devices discussed here. All transport measurements were performed in a He$^3$ Heliox system with a base temperature of 250 mK. We investigated in full detail two monolayer devices showing identical results. The data presented in the main text are taken from one of the devices which has a width of $W$ = 2.5 μm, and three pairs of Hall probes (the longest distance between different pairs of Hall probes in this device is 5.5 μm, see Fig. 4b). The thickness of WS$_2$ is about 26 nm. Data from the second monolayer device, virtually identical to the first one, are shown in the Supplementary Information.

Fully relativistic density functional theory (DFT) calculations were performed using the Vienna ab-initio Simulation Package (VASP) with projector augmented wave (PAW) pseudopotentials under the generalized gradient approximation (GGA)[51,52]. The graphene lattice constant was always kept at 2.46 Å. Ionic relaxations were performed for WS$_2$ in 1×1 unit cells with lattice constants fixed to three different rational multiples (4/3, 5/4, and 9/7) of that of graphene. Supercells with different moiré periodicities were then constructed by repeating these unit cells correspondingly and by aligning the lattice vectors of WS$_2$ with that of graphene without further ionic relaxation. The separation between graphene and WS$_2$ was fixed to a number of different values ranging from 2.3 Å to 3.3 Å for each supercell. A Monkhorst-Pack $k$-point mesh[53] of 6×6×1 was used for the 4×4 and 5×5 supercells (in terms of graphene lattice constant), and that of 3×3×1 was used for the 9×9 supercell. The plane wave energy cutoff was set to 400 eV in all calculations.

**Acknowledgements**

We gratefully acknowledge technical assistance from A. Ferreira. ZW, DK and AFM also gratefully acknowledge financial support from the Swiss National Science Foundation, the NCCR QSIT, and the EU Graphene Flagship Project. AHM and HC were supported by ONR-N00014-14-1-0330 and Welch Foundation grant TBF1473. HC and AHM acknowledge the Texas Advanced Computing Center (TACC) at The University of Texas at Austin for providing HPC resources for the *ab-initio* calculations in this work. HC thanks Y. Araki, X. Li and A. DaSilva for valuable discussions.

**Author contributions**

ZW fabricated the devices and performed the measurements with the collaboration of DK, under the supervision of AFM. HB provided the $WS_2$ crystals used to realize the devices. ZW, DK and AFM analyzed the data. HC performed the numerical calculations under the guidance of AHM. All authors discussed and interpreted the results, and contributed to writing the paper.

**Competing financial interests** The authors declare no competing financial interests.

Correspondence and requests for materials should be addressed to A.F.M. (Alberto.Morpurgo@unige.ch).

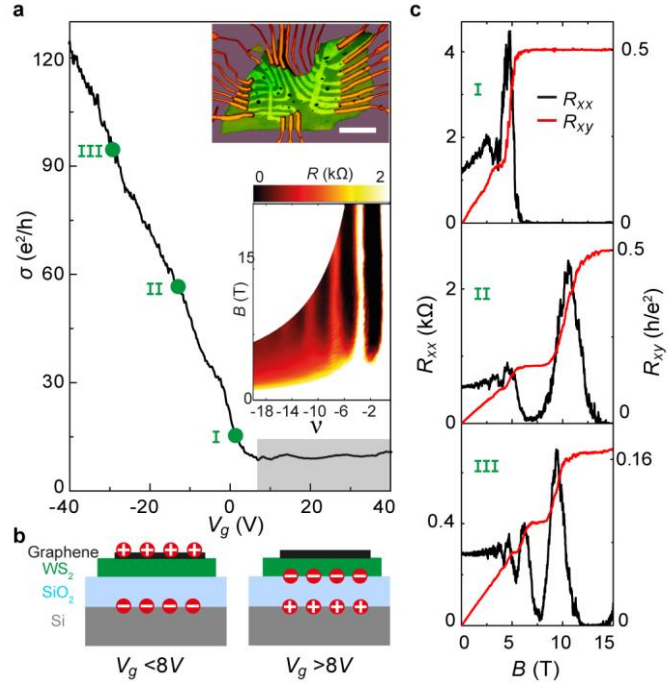

**Fig. 1 | Basic electrical transport properties of graphene-on-WS$_2$ heterostructure at 250 mK. a,** The conductivity ($\sigma$) of the device as a function of $V_g$, exhibiting a typical linear dependence at $V_g$ below ~8 V above which (shadow area) it saturates. Green dots mark the ranges of $V_g$ (I: 0 V to 5 V, II: -10 V to -15 V, and III: -25 V to -30 V) where the experimental results presented in Figs. 2 and 3 are obtained. The upper inset displays the false-colored optical image of a typical device with a Hall-bar geometry (the scale bar is 10 μm). The lower inset exhibits the Shubnikov-de Hass oscillation of the longitudinal resistance ($R_{xx}$) originating from the characteristic half-integer quantum-Hall effect. The $R_{xx}$ minima (black strips) at a half-integer sequence of filling factor $|\nu| = |nh/eB| = 4 \times (N+1/2)$ are clearly visible ($N$: integer). **b,** Schematic cross-section of the device illustrating a vertical profile of the charge accumulation at $V_g$ below and above ~8 V (on the left and the right, respectively). Due to the presence of the n-doped WS$_2$ substrate, charges are accumulated in graphene only at $V_g$ lower than ~8 V applied to the back-gate (a highly doped silicon substrate covered with 280-nm-thick SiO$_2$ layer). **c,** Fully developed half-integer quantum-Hall effect with vanishing $R_{xx}$ (black curve) and quantized $R_{xy}$ (red curve) is observed at different values of $V_g = 0$ V, -11.5 V, and -28 V (from top to bottom) within the region I, II, and III (indicated in **a**).

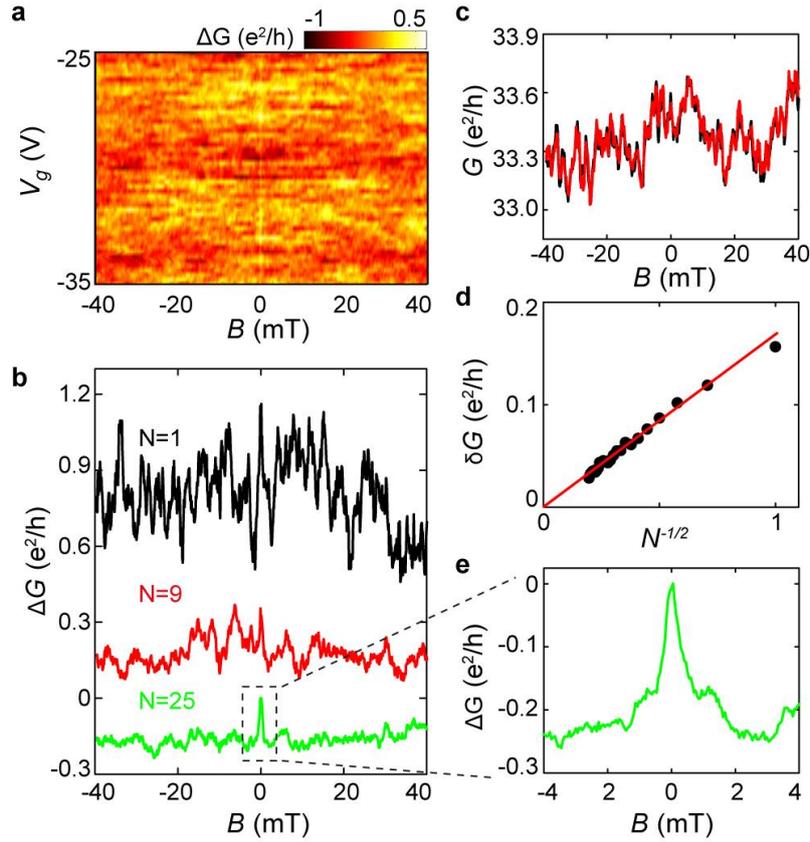

**Fig. 2 | Ensemble averaging of the magneto-conductance at 250 mK. a,** Color-coded MC, $\Delta G(B)$, as a function of $V_g$, with the background conductance slowly varying in $V_g$ (shown in Fig. 1a) subtracted. The large background conductance fluctuations originating from phase-coherent interference of electron waves are apparent, and nearly completely obscure the effect of WAL around $B = 0$ T. **b,** Evolution of the averaged MC upon increasing the number $N$ of uncorrelated MC traces used to calculate the average ($N = 1, 9, 25$; curves offset for clarity): the conductance peak at zero $B$ associated to WAL becomes apparent for sufficiently large values of $N$. **c,** Two MC traces measured at $V_g = -25$ V (red and black curves) demonstrating the excellent reproducibility of the conductance fluctuations. **d,** After averaging over $N$ different curves, the root-mean-square amplitude of the conductance fluctuations ($\delta G$) decreases proportionally to $N^{1/2}$ as expected for a proper ensemble-averaging process. **e,** Zoomed-in view of the ensemble-averaged MC (for $N=25$), which clearly exhibits a sharp conductance peak at $B = 0$ T.

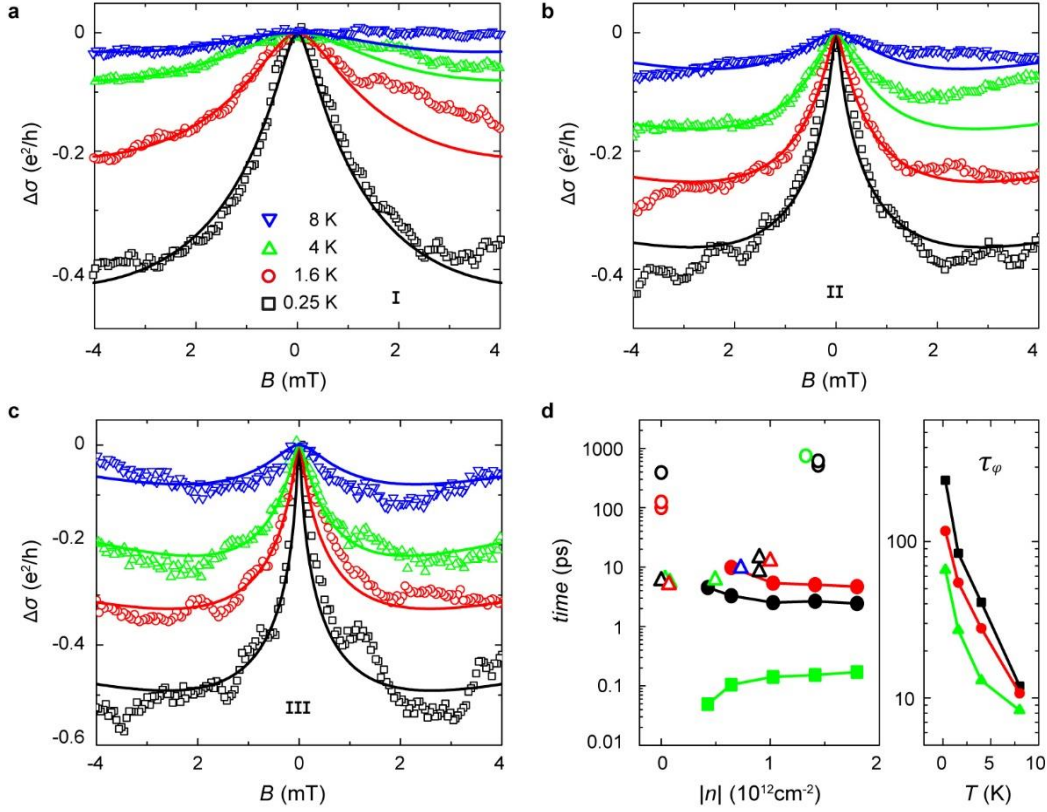

**Fig. 3 | Low-temperature weak anti-localization in graphene-on-WS$_2$. a-c,** Ensemble-averaged MC curves (symbols) obtained from measurements performed in different ranges of $V_g$ (I, II, and III, respectively, indicated in Fig. 1a), at several different temperatures below 8 K. The square magneto-conductance $\Delta\sigma = \sigma(B\neq 0) - \sigma(B=0)$ clearly exhibits a peak at zero $B$ in all $V_g$-ranges, whose height decreases as temperature is increased from 250 mK to 8 K, the expected behavior of WAL due to SOI. Solid lines show the best fits to Eq. (1) in the main text. **d,** Left panel: carrier density dependence of the relevant characteristic times. The filled squares represent the elastic scattering time ($\tau$) estimated from the conductivity of our device at zero $B$; the filled black (red) circles represent the spin-relaxation time ($\tau_{so}$) extracted from the analysis of WAL (non-local spin-Hall effect; see Supplementary Information). For comparison, open up-triangles represent the values of intervalley scattering time ($\tau_{iv}$) reported in the literature, and extracted from the analysis of weak-localization measured in device similar to ours on different substrates, such as SiO$_2$ (black[27] and red[29]), hBN (green[30]), and GaAs (blue[31]). Open circles represent $\tau_{so}$ obtained from spin-valve studies on pristine graphene on SiO$_2$ (black[37] and red[38]) and hBN (green[40]). Right panel: temperature dependence of the phase-coherence time ($\tau_\phi$) of electrons in graphene-on-WS$_2$ extracted from the analysis of WAL performed in this work, for different gate-voltage ranges (I, II, and III from bottom to top). The data clearly exhibit an increase in $\tau_\phi$ with lowering temperature.

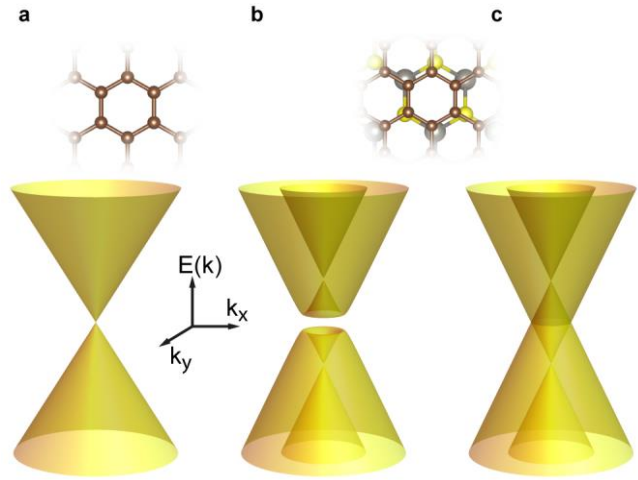

**Fig. 4 | Low-energy band structure of graphene-on-WS$_2$ near the K/K' point. a,** The usual "Dirac cone" low-energy dispersion relation for spin-degenerate charge carriers in graphene close to the K/K' point. **b-c,** At an interface with a WS$_2$ substrate, the dispersion relation is modified by the effect of the induced spin-orbit interaction. *ab-initio* calculations show that the low-energy Hamiltonian in Eq. (2) accurately describes the modifications to the band-structure of graphene. Two spin-orbit interaction terms, with coupling constant $\lambda$ and $\lambda_R$, are induced by interfacial interactions. Our calculations indicate that $\lambda \sim 5$ meV and $\lambda_R \sim 1$ meV. With these values the dispersion relation of electrons becomes the one shown in **b** that, at charge neutrality, corresponds to the band-structure of an insulator (with non-trivial topological properties; see Supplementary Information). The size of the gap is determined by the value of $\lambda_R$ (as long as $\lambda \gg \lambda_R$). Since the gap is likely small in our devices as compared to electrostatic potential fluctuations, $\lambda_R$ can be neglected in a first approximation, in which case the dispersion relation becomes the one shown in **c**.

# References

1. Kane, C. L. & Mele, E. J. Z(2) topological order and the quantum spin Hall effect. *Phys. Rev. Lett.* **95**, 146802 (2005).
2. Kane, C. L. & Mele, E. J. Quantum spin Hall effect in graphene. *Phys. Rev. Lett.* **95**, 226801 (2005).
3. Huertas-Hernando, D., Guinea, F. & Brataas, A. Spin-orbit coupling in curved graphene, fullerenes, nanotubes, and nanotube caps. *Phys. Rev. B* **74**, 155426 (2006).
4. Min, H. *et al.* Intrinsic and Rashba spin-orbit interactions in graphene sheets. *Phys. Rev. B* **74**, 165310 (2006).
5. Konschuh, S., Gmitra, M. & Fabian, J. Tight-binding theory of the spin-orbit coupling in graphene. *Phys. Rev. B* **82**, 245412 (2010).
6. Castro Neto, A. H. & Guinea, F. Impurity-Induced Spin-Orbit Coupling in Graphene. *Phys. Rev. Lett.* **103**, 026804 (2009).
7. Weeks, C., Hu, J., Alicea, J., Franz, M. & Wu, R. Q. Engineering a Robust Quantum Spin Hall State in Graphene via Adatom Deposition. *Phys. Rev. X* **1**, 021001 (2011).
8. Han, W., Kawakami, R. K., Gmitra, M. & Fabian, J. Graphene spintronics. *Nat. Nanotechnol.* **9**, 794-807 (2014).
9. Balakrishnan, J., Koon, G. K. W., Jaiswal, M., Neto, A. H. C. & Ozyilmaz, B. Colossal enhancement of spin-orbit coupling in weakly hydrogenated graphene. *Nat. Phys.* **9**, 284-287 (2013).
10. Avsar, A. *et al.* Spin-orbit proximity effect in graphene. *Nat. Commun.* **5**, 4875 (2014).
11. Calleja, F. *et al.* Spatial variation of a giant spin-orbit effect induces electron confinement in graphene on Pb islands. *Nat. Phys.* **11**, 43-47 (2015).
12. Dean, C. R. *et al.* Hofstadter's butterfly and the fractal quantum Hall effect in moire superlattices. *Nature* **497**, 598-602 (2013).
13. Hunt, B. *et al.* Massive Dirac Fermions and Hofstadter Butterfly in a van der Waals Heterostructure. *Science* **340**, 1427-1430 (2013).
14. Ponomarenko, L. A. *et al.* Cloning of Dirac fermions in graphene superlattices. *Nature* **497**, 594-597 (2013).
15. Dean, C. R. *et al.* Boron nitride substrates for high-quality graphene electronics. *Nat. Nanotechnol.* **5**, 722-726 (2010).
16. Kretinin, A. V. *et al.* Electronic Properties of Graphene Encapsulated with Different Two-Dimensional Atomic Crystals. *Nano Lett.* **14**, 3270-3276 (2014).
17. Wang, Q. H., Kalantar-Zadeh, K., Kis, A., Coleman, J. N. & Strano, M. S. Electronics and optoelectronics of two-dimensional transition metal dichalcogenides. *Nat. Nanotechnol.* **7**, 699-712 (2012).
18. Zhu, Z. Y., Cheng, Y. C. & Schwingenschlogl, U. Giant spin-orbit-induced spin splitting in two-dimensional transition-metal dichalcogenide semiconductors. *Phys. Rev. B* **84**, 153402 (2011).
19. Xiao, D., Liu, G. B., Feng, W. X., Xu, X. D. & Yao, W. Coupled Spin and Valley Physics in Monolayers of $MoS_2$ and Other Group-VI Dichalcogenides. *Phys. Rev. Lett.* **108**, 196802 (2012).
20. Kosmider, K., Gonzalez, J. W. & Fernandez-Rossier, J. Large spin splitting in the conduction band of transition metal dichalcogenide monolayers. *Phys. Rev. B* **88**, 245436 (2013).
21. Larentis, S. *et al.* Band Offset and Negative Compressibility in Graphene-$MoS_2$ Heterostructures. *Nano Lett.* **14**, 2039-2045 (2014).
22. Braga, D., Gutiérrez Lezama, I., Berger, H. & Morpurgo, A. F. Quantitative Determination of the Band Gap of $WS_2$ with Ambipolar Ionic Liquid-Gated Transistors. *Nano Lett.* **12**, 5218-5223 (2012).
23. Castro Neto, A. H., Guinea, F., Peres, N. M. R., Novoselov, K. S. & Geim, A. K. The electronic properties of graphene. *Rev. Mod. Phys.* **81**, 109-162 (2009).
24. Hikami, S., Larkin, A. I. & Nagaoka, Y. Spin-Orbit Interaction and Magnetoresistance in the Two Dimensional Random System. *Prog. Theor. Phys.* **63**, 707-710 (1980).

| | |
|---|---|
| 25 | Beenakker, C. W. J. & van Houten, H. in *Solid State Physics* Vol. 44 (eds H. Ehrenreich & D. Turnbull) 1-228 (Academic Press, New York, 1991). |
| 26 | Lee, P. A. & Stone, A. D. Universal Conductance Fluctuations in Metals. *Phys. Rev. Lett.* **55**, 1622-1625 (1985). |
| 27 | Tikhonenko, F. V., Horsell, D. W., Gorbachev, R. V. & Savchenko, A. K. Weak localization in graphene flakes. *Phys. Rev. Lett.* **100**, 056802 (2008). |
| 28 | Tikhonenko, F. V., Kozikov, A. A., Savchenko, A. K. & Gorbachev, R. V. Transition between Electron Localization and Antilocalization in Graphene. *Phys. Rev. Lett.* **103**, 226801 (2009). |
| 29 | Lundeberg, M. B. & Folk, J. A. Rippled Graphene in an In-Plane Magnetic Field: Effects of a Random Vector Potential. *Phys. Rev. Lett.* **105**, 146804 (2010). |
| 30 | Couto, N. J. G. *et al.* Random Strain Fluctuations as Dominant Disorder Source for High-Quality On-Substrate Graphene Devices. *Phys. Rev. X* **4**, 041019 (2014). |
| 31 | Woszczyna, M., Friedemann, M., Pierz, K., Weimann, T. & Ahlers, F. J. Magneto-transport properties of exfoliated graphene on GaAs. *J. Appl. Phys.* **110**, 043712 (2011). |
| 32 | Suzuura, H. & Ando, T. Crossover from Symplectic to Orthogonal Class in a Two-Dimensional Honeycomb Lattice. *Phys. Rev. Lett.* **89**, 266603 (2002). |
| 33 | McCann, E. *et al.* Weak-Localization Magnetoresistance and Valley Symmetry in Graphene. *Phys. Rev. Lett.* **97**, 146805 (2006). |
| 34 | Morpurgo, A. F. & Guinea, F. Intervalley Scattering, Long-Range Disorder, and Effective Time-Reversal Symmetry Breaking in Graphene. *Phys. Rev. Lett.* **97**, 196804 (2006). |
| 35 | McCann, E. & Fal'ko, V. I. $z \to -z$ Symmetry of Spin-Orbit Coupling and Weak Localization in Graphene. *Phys. Rev. Lett.* **108**, 166606 (2012). |
| 36 | Abanin, D. A., Shytov, A. V., Levitov, L. S. & Halperin, B. I. Nonlocal charge transport mediated by spin diffusion in the spin Hall effect regime. *Phys. Rev. B* **79**, 035304 (2009). |
| 37 | Tombros, N., Jozsa, C., Popinciuc, M., Jonkman, H. T. & van Wees, B. J. Electronic spin transport and spin precession in single graphene layers at room temperature. *Nature* **448**, 571 (2007). |
| 38 | Han, W. & Kawakami, R. K. Spin Relaxation in Single-Layer and Bilayer Graphene. *Phys. Rev. Lett.* **107**, 047207 (2011). |
| 39 | Zomer, P. J., Guimaraes, M. H. D., Tombros, N. & van Wees, B. J. Long-distance spin transport in high-mobility graphene on hexagonal boron nitride. *Phys. Rev. B* **86**, 161416 (2012). |
| 40 | Guimaraes, M. H. D. *et al.* Controlling Spin Relaxation in Hexagonal BN-Encapsulated Graphene with a Transverse Electric Field. *Phys. Rev. Lett.* **113**, 086602 (2014). |
| 41 | Elliott, R. J. Theory of the Effect of Spin-Orbit Coupling on Magnetic Resonance in Some Semiconductors. *Phys. Rev.* **96**, 266-279 (1954). |
| 42 | Yafet, Y. in *Solid State Physics* Vol. 14 (eds F. Seitz & D. Turnbull) 1-98 (Academic Press Inc., New York, 1963). |
| 43 | Dyakonov, M. & Perel, V. Spin relaxation of conduction electrons in noncentrosymmetric semiconductors. *Sov. Phys. Solid State* **13**, 3023-3026 (1972). |
| 44 | Tuan, D., Ortmann, F., Soriano, D., Valenzuela, S. O. & Roche, S. Pseudospin-driven spin relaxation mechanism in graphene. *Nat. Phys.* **10**, 857-863 (2014). |
| 45 | Jung, J., Raoux, A., Qiao, Z. H. & MacDonald, A. H. Ab initio theory of moire superlattice bands in layered two-dimensional materials. *Phys. Rev. B* **89**, 205414 (2014). |
| 46 | Haigh, S. J. *et al.* Cross-sectional imaging of individual layers and buried interfaces of graphene-based heterostructures and superlattices. *Nat. Mater.* **11**, 764-767 (2012). |
| 47 | Roth, A. *et al.* Nonlocal Transport in the Quantum Spin Hall State. *Science* **325**, 294-297 (2009). |
| 48 | Wang, L. *et al.* One-Dimensional Electrical Contact to a Two-Dimensional Material. *Science* **342**, 614-617 (2013). |
| 49 | Goossens, A. M. *et al.* Mechanical cleaning of graphene. *Appl. Phys. Lett.* **100**, 073110 (2012). |
| 50 | *Topological Insulators* in *Contemporary Concepts of Condensed Matter Science* Vol. 6 (eds Franz Marcel & Molenkamp Laurens) 1-324 (Elsevier, 2013). |

51  Kresse, G. & Hafner, J. Ab initio molecular dynamics for liquid metals. *Phys. Rev. B* **47**, 558-561 (1993).
52  Kresse, G. & Furthmuller, J. Efficient iterative schemes for ab initio total-energy calculations using a plane-wave basis set. *Phys. Rev. B* **54**, 11169-11186 (1996).
53  Monkhorst, H. J. & Pack, J. D. Special Points for Brillouin-Zone Integrations. *Phys. Rev. B* **13**, 5188-5192 (1976).

# Supplementary Information for

# Strong interface-induced spin-orbit coupling in graphene on $WS_2$

**S1. Saturation of charge density in graphene-on-$WS_2$ at positive $V_g$ occurring concomitantly with the conductivity saturation.**

In the main text, we show that the conductivity ($\sigma$) of graphene-on-$WS_2$ devices saturates for $V_g > \sim 8$ V, because upon changing the gate voltage in this range, charges are accumulated at the $WS_2$/$SiO_2$ interface (where the carrier mobility is low) and not in graphene. Here, we confirm this conclusion by measuring the Hall effect in different gate-voltages ranges, to show that the Hall density ($n$) remains constant in the interval of $V_g$ where $\sigma$ saturates. Figs. S1a-b show the Hall resistance ($R_{Hall}$) as a function of magnetic field ($B$), measured for values of $V_g$ below and above 0 V, respectively. It is apparent that the slope of this curve, which measures the charge density ($n$) accumulated in graphene, changes upon varying $V_g$ below 0 V, and remains unchanged for positive $V_g$. A comparison between Fig. S1c and S1d, in which the carrier density and the conductivity are shown as a function of $V_g$ between -40 V and 40 V, demonstrates that the saturation of $n$ parallels the saturation of $\sigma$, as expected.

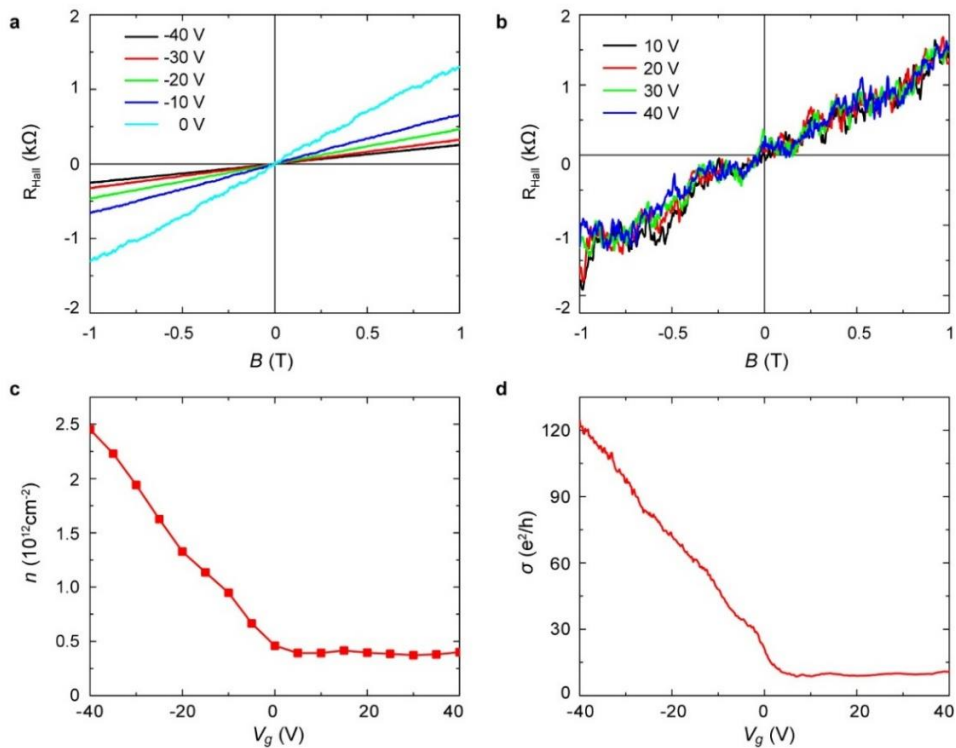

**Fig. S1 | Classical Hall effect and charge density as a function of back-gate voltage. a-b,** Hall resistance ($R_{Hall}$) as a function of magnetic field ($B$) measured at different values of gate-voltage, below and above 0 V. The slope, which measures the charge density accumulated in graphene, remains unchanged when the device is positively gate-biased. **c,** Gate-voltage ($V_g$) dependence of the Hall density ($n$) extracted from the measurement of the Hall resistance. **d,** Conductivity as a function of $V_g$, exhibiting a $V_g$ dependence paralleling that of $n$ shown in panel **c**.

## S2. Weak anti-localization effect measured in different devices.

As discussed in detail in the main text, the observation of weak anti-localization (WAL) at low temperature unambiguously demonstrates the presence of strong SOI in graphene-on-$WS_2$. To illustrate the reproducibility of our observations, here we show data from a different device measured at the lowest temperature of our cryostat ($T$ = 250 mK), which are virtually identical to those obtained from the device discussed in the main text. Fig. S2a shows that the conductivity varies linearly with $V_g$ only for negative gate voltages, and the inset illustrates the $V_g$-dependence of $n$. The quantum Hall effect data in Fig. S2b clearly confirms that holes in the device behave as Dirac fermions[1]. Since the device dimension is similar to that of the device discussed in the main text, we reveal WAL by ensemble averaging the measured magnetoconductance around three different values of gate voltage; the results of the averaging are shown in Fig. S2c. It is apparent that a sharp conductance peak at zero magnetic field appears in all gate-voltage ranges explored, with an amplitude as large as ~$0.5e^2/h$ at the largest negative gate-voltages. The data are well fit by Eq. (1) of the main text[2] (see the continuous lines in Fig. S2c); the uncertainty in the fitting parameters originates from the effect of the residual random conductance fluctuations remaining after the averaging process. The characteristic times extracted from the fitting are shown in Fig. S2d, and exhibit values and trends virtually identical to those obtained from the device discussed in the main text.

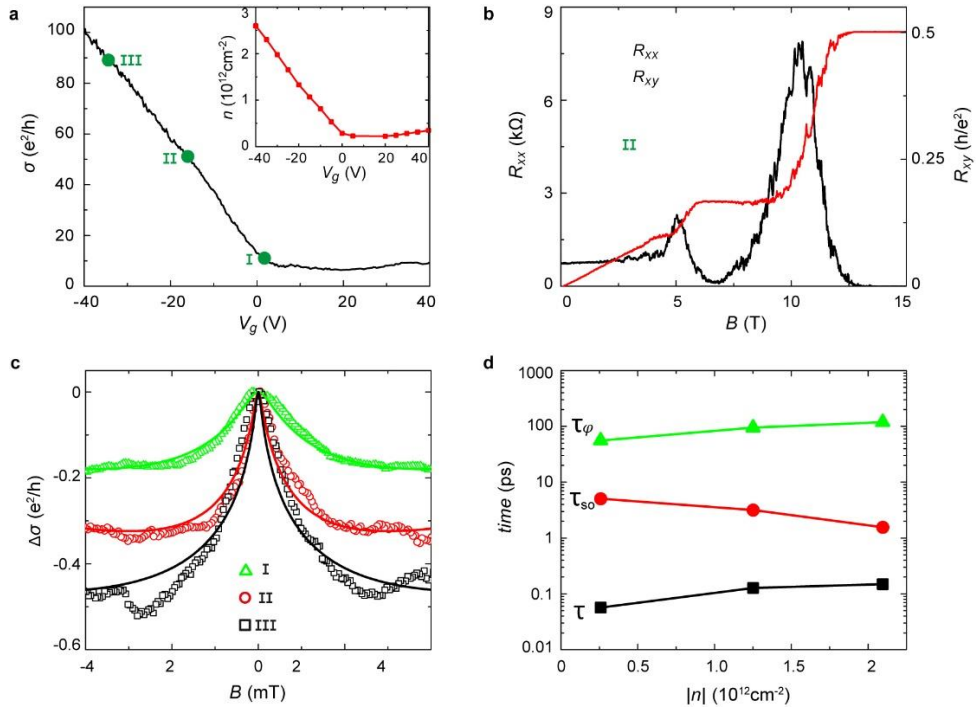

**Fig. S2 | Basic device characterization and observation of weak anti-localization in device 2 at 250 mK. a-b,** The basic characterization of device 2 shows a behavior virtually identical to that of the device discussed in the main text. The measured gate-voltage ($V_g$) dependence of the conductivity ($\sigma$) saturates at positive $V_g$ (**a**; the inset further shows the concomitant saturation of $n$), and clear half-integer quantum-Hall effect is found for negative $V_g$ (see panel **b**). **c,** Ensemble-averaged magnetoconductance in the three different gate-voltage ranges (I: 5 ~ 0 V, II: -15 ~ -20 V, and III: -30 ~ -35 V) indicated in panel **a**. The sharp conductance peak at zero $B$ –the characteristic feature of weak anti-localization– is clearly visible in all cases. **d,** Density dependence of the estimated phase-coherence time ($\tau_\phi$; green up-triangles), spin-relaxation time ($\tau_{so}$; red circles), and momentum relaxation time ($\tau$; black squares).

## S3. Non-local resistance as a signature of SOI.

We note that demonstrating the presence of strong SOI by measuring WAL effect at low temperatures, an entirely established method, has not been achieved previously in graphene. Related earlier work, including a recent one (further discussed in the section 5) also focusing on graphene-on-WS$_2$ [3], has relied on another transport phenomenon, namely the measurement of non-local resistance ($R_{NL}$) generated through the combination of spin-Hall and inverse spin-Hall effect (often referred to as the non-local spin-Hall effect)[4]. This method works when the contribution to the non-local resistance due to this phenomenon is larger than the Ohmic contribution (which can be estimated to be $R_{ohm} = \rho e^{-\pi L/W}$, $\rho$ is the resistivity and $L/W$ is the device aspect ratio). The non-local signal due to the spin-Hall and inverse spin Hall effect decays exponentially away from the contacts used to inject current on a characteristic scale determined by the spin-relaxation length $\lambda_{so} = (D\tau_{so})^{1/2}$. Although our devices are not intentionally designed to optimally perform non-local measurements, the multi-terminal Hall bar geometry nevertheless allows us to probe non-local signals. The results of these non-local measurements for $V_g < 0$ V (i.e., away from the $V_g$-region where the conductivity of graphene saturates) are shown in Fig. S3. It is apparent that $R_{NL}$ in this regime is larger than $R_{Ohm}$ (by approximately a factor of 2 to 3, depending on $V_g$, Fig. S3a) and that it decays very rapidly with increasing $L$, the distance between the contacts used to inject current and those used to detect voltage (Fig. S3b). The data are compatible with an exponential decay (black line in Fig. S3b) and using the formula $R_{NL} = \gamma^2 \rho W/2\lambda_{so} e^{-L/\lambda_{so}}$ for the expected behavior of the non-local signal due to the spin-Hall effect ($\gamma$ is the spin-Hall coefficient), we estimate values of $\tau_{so}$ (see the filled red dots in Fig. 3d of the main text) that are very close to those inferred from the analysis of WAL (the factor 2 difference is certainly compatible with the errors associated to the non-ideal device geometry for the analysis of non-local effect and with the precision of the WAL analysis; as mentioned in the main text, such level of uncertainty in $\tau_{so}$ does not affect the conclusions of our study).

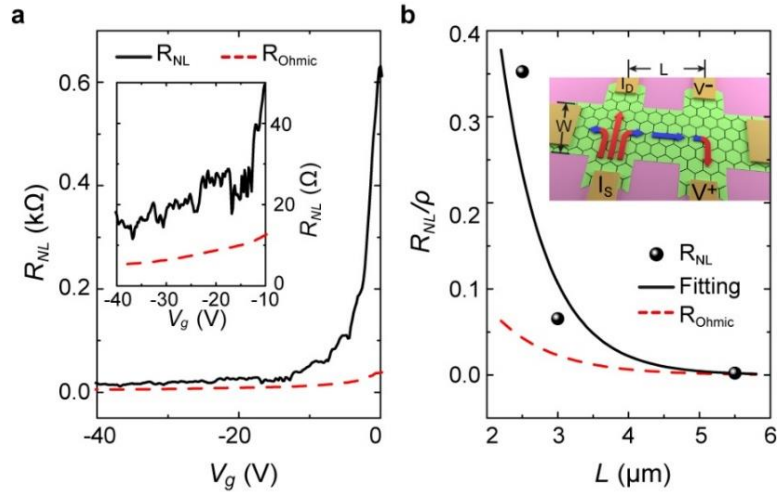

**Fig. S3 | Non-local resistance measurement at 1.6 K. a,** $V_g$-dependence of the non-local resistance $R_{NL}$ (black line) compared to the contribution from Ohmic transport (red broken line), $R_{ohm} = \rho e^{-\pi L/W}$ (with $L/W = 1.2$ in the present case). $R_{NL}$ is larger than $R_{ohm}$ in the entire $V_g$ range explored. **b,** At a fixed $V_g = -27$ V, the measured $R_{NL}$ (black dots) exhibits a very rapid decay upon increasing $L$, consistent with an exponential behavior (quantitatively not consistent with the behavior expected for the Ohmic signal, represented by the red dashed line). As illustrated in the inset, $R_{NL} = V_{NL}/I_c$ is obtained by injecting a charge current $I_c$ (red arrow) from source to drain contacts (denoted by $I_S$ and $I_D$, respectively), and measuring the non-local voltage $V_{NL} = V^+ - V^-$ between two probes (denoted correspondingly) separated from the injecting contacts by a distance $L$.

## S4. Electronic Structure of Graphene on WS$_2$.

The role of spin-orbit coupling in the electronic structure of graphene on WS$_2$ can be addressed theoretically using the tools of electronic structure theory. This approach is complicated by the lattice constant mismatch between the two materials, and by the fact that the separation between layers may not be reliably predicted on the basis of purely theoretical considerations. To make progress we have performed electronic structure calculations (details are given in the Methods section) for super cells with three different ratios between the lattice constants of WS$_2$ and graphene, and for a wide range of layer separations. We find that in all cases a set of states whose wavefunctions that are strongly peaked in the graphene layer appear inside the WS$_2$ gap. The electronic structure of these states is always well described by an Hamiltonian in which the Dirac continuum model of isolated graphene is supplemented by three momentum and position independent substrate interaction terms,

$$H = H_0 + \Delta/2\, \sigma_z + \lambda/2\, \tau_z s_z + \lambda_R/2(\tau_z \sigma_x s_y - \sigma_y s_x) \quad \text{(S1)}$$

Here $H_0$ is the spin-independent Dirac Hamiltonian of isolated graphene and $\sigma$, $\tau_z$, and $s$ act respectively on its sublattice, valley, and spin degrees of freedom. The first two terms give rise to spin and valley dependent gapped massive Dirac bands which overlap in energy when $\lambda > \Delta$. The third term leads to avoided crossings at finite momentum when bands with different values of $\tau_z s_z$ overlap. Our *ab initio* numerical calculations showed no evidence of moiré periodicity effects like those that produce strong anomalies in electronic properties at four carriers per moiré period in the case of graphene on hexagonal Boron nitride with small relative orientation angles. A typical superlattice band structure is illustrated in Fig. S4. WS$_2$ and graphene share triangular Bravais lattices but WS$_2$ has a larger lattice constant (~3.16 Å) than that of graphene (2.46 Å). Supercells were constructed by choosing three different rational approximants to the lattice constant ratio, 4:3, 5:4, and 9:7. The 9:7 case (Figs. S4c and S4d) requires that calculations be performed for the largest supercells, but closely approximates the experimental lattice constant ratio. A previous study[3] reported on band calculations for the 4:3 lattice constant ratio.

Since $\tau_z s$ and $\sigma_y$ are odd under time reversal, all three substrate interaction terms are time reversal invariant. Similarly all three terms are absent in an isolated graphene sheet because of inversion symmetry, which maps $\tau_z$ to $-\tau_z$ and transforms the $\sigma$ matrices like $\sigma_x$. Although we expect a term in the low energy model Hamiltonian that is proportional to $\tau_z \sigma_z s_z$, which *is* symmetry allowed even in the isolated graphene case[5], this term is evidently too weak for it to be manifested in the microscopic supercell calculations.

In Fig. S5 we plot coupling constants, obtained by fitting Eq. (S1) to *ab initio* supercell bands, as a function of the separation between graphene and WS$_2$ for 4:3, 5:4, and 9:7 commensurability. In each case the microscopic bands are in principle dependent on the position of the origin of the graphene Bravais lattice relative to the origin of the substrate Bravais lattice. We find however that these rigid translations have negligible effect on the graphene bands in practice. This finding supports the view that the spatial variation of the local coordination between the graphene sheet and the substrate does not lead to an important spatial variation in the effective Hamiltonian. Instead the orbitals of interest are influenced only by a spatially averaged effect of the substrate. When $\Delta \neq 0$ this averaging evidently still allows for a distinction between the graphene sheet sublattices but, as illustrated in Fig. S5 this effect is smaller for the experimentally relevant 9:7 lattice constant ratio and also smaller at the physically relevant layer separation. Experimentally the graphene/WS$_2$ separation has not yet been accurately determined, making it difficult to infer realistic values of the coupling parameters from the DFT calculations. Theoretically the weak van der Waals coupling between graphene and WS$_2$ cannot be correctly accounted for by conventional density-functional-theory approximations like LDA or GGA. Several semi-empirical schemes for including dispersion forces in DFT have been proposed in the past decades, but still have limited predictive power. As an educated guess, we expect the graphene/WS$_2$ separation to be around 3Å based on measurements performed for graphene on h-BN substrates[6]. For this separation one can estimate from Fig. S5, using data from the largest supercell calculations, that $\Delta \approx 0$ meV, $\lambda \approx 5$ meV, and $\lambda_R \approx 1$ meV.

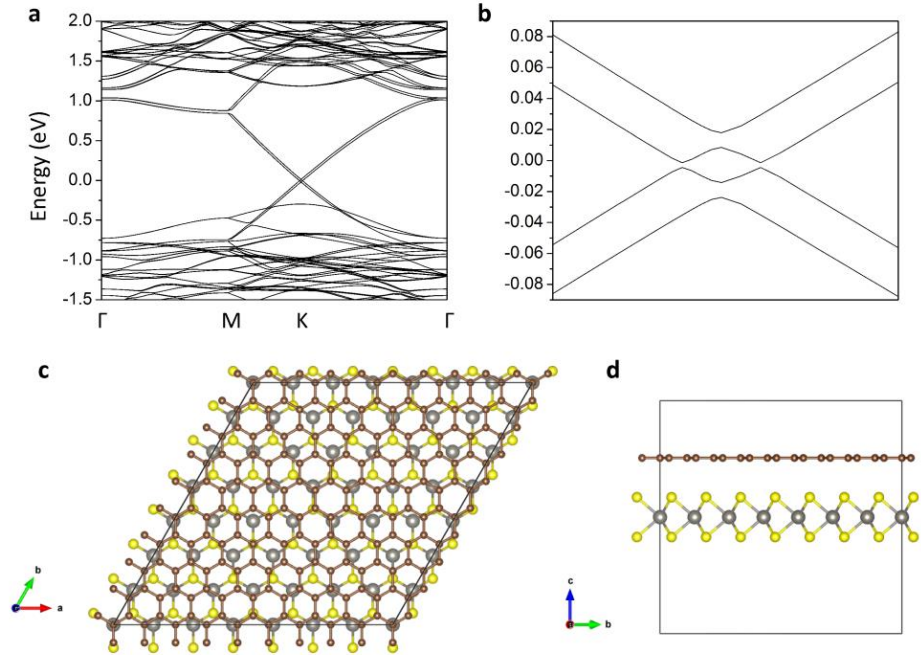

**Fig. S4 Band structure of graphene/WS$_2$ for short period commensurate moiré supercell. a,** band structure of a supercell in which 4×4-repeated monolayer WS$_2$ is lattice matched to 5×5-repeated graphene. In this calculation the separation between the layers was fixed at 2.45 Å. Calculations using a bilayer WS$_2$ substrate which preserves inversion symmetry have also been performed with qualitatively similar band structures for the graphene states inside the WS$_2$ band gap. The momentum labels refer to high-symmetry points in the supercell Brillouin-zone. **b,** zoom view of the low-energy graphene-like bands inside the WS$_2$ band gap. **c,** top and **d,** side views of the moiré supercell for the 9:7 lattice constant ratio case.

Within a single valley, Hamiltonian Eq. S1 has the same form as the low-energy Hamiltonian in Ref. 7 that describes the quantum anomalous Hall (QAH) insulator phase of graphene under both a Zeeman field and Rashba spin-orbit coupling. The difference between Eq. S1 and the Hamiltonian in Ref. 7 is that the Zeeman-like field in the present case, which is supplied by the spin-orbit coupling term proportional to $\lambda$, has opposite signs in the two valleys as required by time-reversal symmetry. Extrapolating from the ideas of Kane and Mele[5], one expects to obtain a topological insulator by combining two mutually time-reversed copies of a QAH insulator. To verify that Eq. S1 indeed describes a topological insulator, we follow Ref. 8 by calculating the $Z_2$ topological invariant using

$$Z_2 = \frac{1}{2\pi}\left[\oint_{\partial B^+} d\mathbf{k}\cdot\mathbf{A}(\mathbf{k}) - \int_{B^+} d^2k\,\Omega_z(\mathbf{k})\right] \mod 2 \tag{S2}$$

where $B^+$ and $\partial B^+$ are respectively a half Brillouin zone and its boundary, and $\mathbf{A}(\mathbf{k})$ and $\Omega_z(\mathbf{k})$ are respectively the Berry connection and the Berry curvature summed over filled bands. $Z_2=1$ for a topological insulator and $Z_2=0$ for a trivial insulator. In the present case the half Brillouin zone can be identified with an area around a single valley and a boundary that is sufficiently far from the centre of the valley. We calculated the $Z_2$ number for $\lambda$ much larger than $\lambda_R$ and $\Delta$ and found that $Z_2$ is always 1. Therefore in the ideal situation in which the low energy physics of the graphene/WS$_2$ system is completely determined by Eq. S1 the system should be a topological insulator. It is worth noting is that in the original Kane-Mele model the topologically nontrivial gap is opened by the intrinsic spin-orbit coupling of graphene which preserves inversion symmetry, whereas in the present model the gap is opened by two spin-orbit coupling terms, both of which break inversion symmetry and are absent in freestanding graphene.

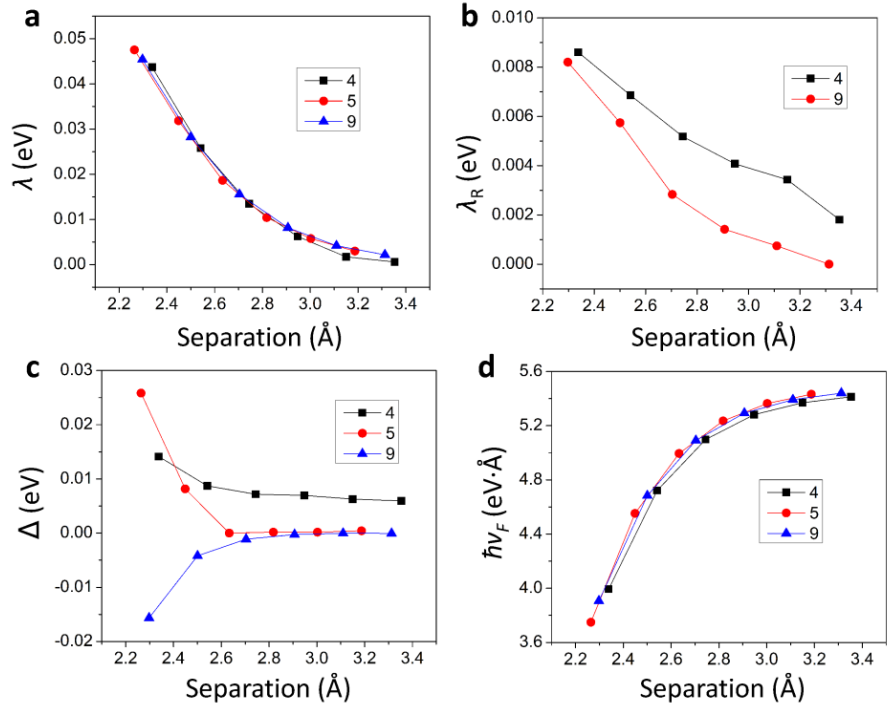

**Fig. S5 | Coupling constants in the low-energy effective model of graphene on WS$_2$ obtained from DFT. a-d,** plot the dependence of $\lambda$, $\lambda_R$, $\Delta$, and $\hbar v_F$ on the separation between graphene and WS$_2$ for three different commensurate moiré supercells. In **b** the case of the 5×5 supercell is not plotted since the splitting due to $\lambda_R$ is too small in this case to enable a reliable estimate. Unlike $\lambda$ and $\hbar v_F$, whose values collapse on the same curves for different supercell sizes, those of $\lambda_R$ and $\Delta$ are apparently different for different supercells. The values of these terms are evidently sensitive to small changes in the average local coordination between graphene and WS$_2$. Because the lattice constant ratio in the 9:7 commensurate supercell calculations is very close to the experimental ratio, the parameters inferred from these calculations should provide a good representation of the experimental situation.

## S5. Discussions of a recent study of SOI in graphene-on-WS$_2$.

In previous work by A. Avsar *et al.*[3] the role of spin-orbit interaction in graphene/WS$_2$ was studied by performing non-local transport measurements. (See Section 3.) Although the present study and the earlier work agree in concluding that SOI is induced in graphene by the WS$_2$ substrate, there exist several important differences. In Ref. 3, it is claimed that charge scattering by sulfur vacancies in the WS$_2$ substrate is the origin of strong SOI, not interfacial interactions with the substrate. *ab initio* calculations, used to support its claim, were also reported, showing no signature of SOI on the band structure of graphene-on-WS$_2$. The following comments are intended to shed light on the origin of these differences.

1. In Ref. 3, experimental evidence for an enhanced SOI is found only for positive $V_g$, in the range where the measured conductance saturates, *i.e.* in the gate voltage range over which the transport properties of graphene cannot be tuned by the back gate. A. Avsar *et al.* also stated that there is no signature of WAL for negative $V_g$ (see Supplementary Figure 10 in Ref. 3 and the discussion related to it), a finding that was argued to support the consistency of their results. However, as clearly shown in our study, WAL in graphene-on-WS$_2$ devices is present throughout the investigated range for which the Fermi level is in the valence band (negative $V_g$). Our results also show that revealing the WAL signal requires a careful analysis. In particular, it is necessary to first eliminate the effect of random conductance fluctuations. In addition, in a high-mobility device (mobility of a few tens of thousands cm$^2$V$^{-1}$S$^{-1}$), WAL manifests itself in a signal visible only at very small magnetic field, of order a few milli-Tesla. Neither ensemble averaging nor measurements focusing on an appropriately small magnetic field range were shown in Ref. 3.

2. In Ref. 3, the conductance saturation observed for positive gate voltage is attributed to the presence of S vacancies in WS$_2$ close to graphene. Specifically, upon increasing the carrier density, the Fermi energy in graphene was claimed to align to defect states originating from the Sulfur vacancies. SOI was then claimed to result from interaction of carriers in graphene with vacancies in WS$_2$. We note that, because of simple electrostatics, when charges can be accumulated in WS$_2$, they are accumulated at the interface closer to the gate, since this is the configuration of minimum electrostatic energy. Considering the accumulation of charges in WS$_2$ at the interface with graphene, therefore, does not seem a consistent interpretation and, consequently, an appropriate description of the origin of the induced SOI.

3. There also exist differences in the theoretical calculations. Even though the study of Ref. 3 and our work use the same methods to calculate the band structure of graphene-on-WS$_2$, the results are notably different. In Ref. 3, the calculation showed no signature of SOI in the band structure while in our study, SOI is clearly visible. As illustrated by Fig. S4, the absence of apparent SOI effects in Ref. 3 might be due to the layer separations at which those calculations were performed.

**References**

1. Castro Neto, A. H., Guinea, F., Peres, N. M. R., Novoselov, K. S. & Geim, A. K. The electronic properties of graphene. *Rev. Mod. Phys.* **81**, 109-162 (2009).
2. McCann, E. & Fal'ko, V. I. $z \rightarrow -z$ Symmetry of Spin-Orbit Coupling and Weak Localization in Graphene. *Phys. Rev. Lett.* **108**, 166606 (2012).
3. Avsar, A. *et al.* Spin-orbit proximity effect in graphene. *Nat. Commun.* **5**, 4875 (2014).
4. Abanin, D. A., Shytov, A. V., Levitov, L. S. & Halperin, B. I. Nonlocal charge transport mediated by spin diffusion in the spin Hall effect regime. *Phys. Rev. B* **79**, 035304 (2009).
5. Kane, C. L. & Mele, E. J. Quantum spin Hall effect in graphene. *Phys. Rev. Lett.* **95**, 226801 (2005).
6. Haigh, S. J. *et al.* Cross-sectional imaging of individual layers and buried interfaces of graphene-based heterostructures and superlattices. *Nat. Mater.* **11**, 764-767 (2012).
7. Tse, W.-K., Qiao, Z., Yao, Y., MacDonald, A. H. & Niu, Q. Quantum anomalous Hall effect in single-layer and bilayer graphene. *Phys. Rev. B* **83**, 155447 (2011).
8. Fu, L. & Kane, C. L. Time reversal polarization and a $Z_2$ adiabatic spin pump. *Phys. Rev. B* **74**, 195312 (2006).